\documentclass[aps,reprint,showpacs,showkeys,superscriptaddress,longbibliography]{revtex4-1}
\usepackage{graphicx}
\usepackage{comment}
\usepackage{enumitem}
\usepackage{multirow}

\usepackage{array}
\usepackage{array}
\newcolumntype{x}[1]{>{\centering\arraybackslash\hspace{0pt}}p{#1}}
\newcolumntype{L}[1]{>{\raggedright\let\newline\\\arraybackslash\hspace{0pt}}p{#1}}
\setlist[itemize]{noitemsep}

\begin{document}

\newcommand{\TabLevels}{
\begin{table*}
\caption{\label{tab:levels}Examples of interactions among the levels of the university (column label denotes level doing the acting and row label denotes level being acted on) and the change activities that our project  engages in at each level.}
\def\arraystretch{1.5}
\begin{tabular}{p{1.3in} x{1.3in} x{1.3in} x{1.3in} x{1.3in} }
\hline
 & \multicolumn{3}{ c }{Examples of Interactions Across Levels} & External Forces \\
 & \textbf{By Administration} & \textbf{By Departments} & \textbf{By Faculty} & \textbf{By SITAR Team} \\

\textbf{On Administration} & --- & Determining priorities for allocating resources & Grassroots faculty committees & Teaching Quality Framework ($\S$~\ref{subsec:TQF}) \\

\textbf{On Departments} & Setting campus priorities and initiatives & --- &  Voting, governance, and committee work  & Visioning \& Alignment Process ($\S$~\ref{subsec:VAP})\\

\textbf{On Faculty} & Measures of teaching effectiveness & Norms for evaluating teaching & --- & Departmental Action Teams ($\S$~\ref{subsec:DAT})\\
\hline
\end{tabular}
\end{table*}
}

\newcommand{\TabInsights}{
\begin{table}
\caption{\label{tab:insights}Key insights from six change perspectives.}
\begin{tabular}{p{\columnwidth}}
\hline
\vspace{.05in}
\textbf{Scientific Management}
\begin{itemize}
\item Use incentives and rewards to influence behavior.
\end{itemize}
\vspace{.05in}
\textbf{Evolutionary}
\begin{itemize}
\item View the university as a holistic system.
\item Pay attention to external factors.
\end{itemize}
\vspace{.05in}
\textbf{Social Cognition}
\begin{itemize}
\item Attend to the underlying beliefs that guide decision-making.
\item For sustainable educational transformations, focus on promoting second-order changes (i.e, changing rather than preserving underlying structures).
\end{itemize}
\vspace{.05in}
\textbf{Cultural}
\begin{itemize}
\item To create lasting change, focus on shifting the underlying culture of a department.
\item Align change efforts with existing cultural features.
\end{itemize}
\vspace{.05in}
\textbf{Political}
\begin{itemize}
\item Build coalitions to support strategic, collective action.
\item Leverage existing internal power structures.
\end{itemize}
\vspace{.05in}
\textbf{Institutional}
\begin{itemize}
\item Leverage existing external structures that influence universities.
\end{itemize} \\

\hline

\end{tabular}
\end{table}
}

\newcommand{\TabExamples}{
\begin{table*}
\caption{\label{tab:examples}Summary of the ways in which the change perspectives influenced the design of our activities.  Different activities at other institutions will be shaped by the change perspectives in different ways, based on local context.}
\def\arraystretch{1.5}
\setlength{\tabcolsep}{0.5em}
\begin{tabular}{p{0.8in} L{1.9in} L{1.9in} L{1.9in}}
\hline
 & Departmental Action Team (Faculty) & Visioning and Alignment \newline Process (Department) & Teaching Quality Framework (Administration) \\

\hline

Scientific \newline Management & Secure course buyouts and service credit for DAT participants. & Communicate with administration to secure additional resources as needed; Create mechanism to reward actions aligned with shared vision. & Shift incentive structures by creating a Teaching Quality Framework. \\

\hline

Evolutionary & Create standing course coordinator positions as a way to adapt to future curricular changes. & Position the department as an educational leader in the university system. & Be flexible and opportunistic in engaging with campus activities that can lead to the creation of a Framework.  \\

\hline

Social \newline Cognition & Analyze data from Institutional Research office; Create course coordinator positions. & Use surveys and interviews to elicit mental maps. & Redefine campus-wide teaching excellence award criteria; Reframe discussion from  ``retention'' to ``student success'' to shift administrators' understanding. \\

\hline

Cultural & Frame DAT as aligned with SEI; Design course coordinator positions to align with existing culture. & Create a shared vision; Frame V\&A process as aligned with SEI efforts, departmental self-perception as an educational leader, and existing democratic processes. & Frame development of the Framework as aligned with existing concerns over student persistence. \\

\hline

Political & Interview faculty to get buy-in for DAT; Get sanction of chair and teaching committee for DAT creation; Propose course coordinator positions to teaching committee before full faculty. & Start process through conversations with chair and executive committee. & Partner with Faculty Assembly and Provost's Persistence Taskforce; Create a faculty taskforce for developing the Framework. \\

\hline

Institutional & Use prestige of funding source to gain entry into department. Align efforts with campus moves to attend to persistence and student success. & Use prestige of funding source to gain entry into department. Align efforts with campus moves to attend to persistence and student success. & Use prestige of funding source to gain attention of administrators. Align efforts with campus moves to attend to persistence and student success. \\
\hline
\end{tabular}
\end{table*}
}

\title{A Framework for Transforming Departmental Culture to Support Educational Innovation}
\date{\today}


\author{Joel C. Corbo}
\affiliation{Center for STEM Learning, University of Colorado, Boulder, CO 80309}

\author{Daniel L. Reinholz}
\affiliation{Center for STEM Learning, University of Colorado, Boulder, CO 80309}

\author{Melissa H. Dancy}
\affiliation{Department of Physics, University of Colorado, Boulder, CO 80309}

\author{Stanley Deetz}
\affiliation{Graduate School, University of Colorado, Boulder, CO 80309}

\author{Noah Finkelstein}
\affiliation{Department of Physics, University of Colorado, Boulder, CO 80309}

\begin{abstract}
This paper provides a research-based framework for promoting institutional change in higher
education. To date, most educational change efforts have focused on relatively narrow subsets
of the university system (e.g., faculty teaching practices or administrative policies)
and have been largely driven by implicit change logics; both of these features have limited
the success of such efforts at achieving sustained, systemic change.  Drawing from the literature
on organizational and cultural change, our framework encourages change agents to coordinate their
activities across three key levels of the university and to ground their activities in the various
change perspectives that emerge from that literature. We use examples from a change project that
we have been carrying out at a large research university to illustrate how our framework can be
used as a basis for planning and implementing holistic change.
\end{abstract}

\maketitle

\section{Introduction}
\label{sec:intro}

\TabLevels

Improving higher education requires more than the development and dissemination of innovative
teaching practices; it requires fundamental changes in the practices and cultures of universities.
Accordingly, this paper provides a framework for creating and sustaining such changes. We developed
this framework in response to numerous national calls to improve STEM (science, technology,
engineering, and mathematics) education by promoting the adoption of active learning
techniques~\cite{PCAST2012, NRC2013}. Active learning focuses on the construction of knowledge
through individual investigation, discussions, and group work rather than the transmission
of knowledge through lecture~\cite{Freeman2014}. Students in active learning courses outperform
their peers in traditional classrooms and are more likely to persist in STEM~\cite{Kogan2014}.
However, despite many attempts to improve STEM education at the college level, active learning
techniques are still not widely adopted.

Physics education research provides insight into the difficulty of changing educational practices.
Because of numerous professional development opportunities (e.g., the Workshop for New Physics
and Astronomy Faculty~\cite{NFW2015} and the CIRTL network~\cite{CIRTL2015}),
nearly all physics faculty are aware of active learning strategies~\cite{Henderson2009}.
However, about one-fifth of physics faculty never try to use such strategies,
and of those  who do try them, about one-third discontinue use after their initial
attempt~\cite{Henderson2009,Henderson2012}. Hence, active learning is not widely implemented
in physics classrooms despite evidence favoring it and numerous efforts to encourage its use.

Typical approaches to educational transformation, like those above, assume that educational
practices that are sufficiently well developed, packaged, and disseminated will eventually
enjoy broad-scale implementation~\cite{Seymour2002, Henderson2011}.  However, this assumption
ignores deep-rooted institutional structures and cultural norms that complicate educational
transformation. These environmental factors tend to discourage the use of educational
innovations, even for faculty who conceive of teaching and learning in ways compatible with
the findings of education research~\cite{Henderson2007}.  For instance, the lack of robust measures
of teaching effectiveness discourages faculty from investing time in their
teaching~\cite{Henderson2014}. Thus, when change efforts fail to account for the university
as a system, focusing only on individual faculty practices, they are less likely to succeed.

Understanding the university as a system requires a focus on culture.  Culture is a constantly
evolving system of shared beliefs, values, customs, rituals, practices, and artifacts
that the members of an organization use to cope with their world and with one another, and that
are transmitted from generation to generation through learning~\cite{Schein2010}. Thus, practices
are but one part of a larger cultural system, which should be targeted holistically, rather than in
isolation. In the above example, efforts to change measures of teaching effectiveness would benefit
from accounting for various aspects of culture, such as the origin of such measures, how they are
perceived by key actors, \textit{and} their impact on university practices.

Our framework was developed as part of the STEM Institutional Transformation Action Research (SITAR)
Project, a three-year grant-funded project to implement and study institutional change at
a large research university (which we refer to in this paper as the ``target university''), 
with a focus on shifting departmental culture to improve undergraduate education.  Thus, our team 
is acting as both researcher and change agent.  In writing this paper, we hope to support anyone 
at a university trying to make sustainable change in how their institution educates students.  
These change projects may arise due to external pressure, top-down mandates, or grassroots needs.
They may be supported by external funding or volunteer effort.  They may be initiated by faculty,
administrators, or students.  No matter the form of the change effort, this paper takes 
as its starting point that the reader has a desire to create change and is looking for a framework 
to support such work.

In our framework, we conceptualize a university as a multi-leveled, interconnected system, 
and argue that change efforts should target all of these levels in a coordinated fashion 
(see $\S$~\ref{sec:levels}). Synthesizing the organizational change literature, we highlight 
key factors for successful change (see $\S$~\ref{sec:theories}). We then illustrate our framework
through examples from our change efforts at the target university (see $\S$~\ref{sec:examples}).
Because the purpose of this paper is to provide a framework, not to evaluate the efforts themselves,
we leave more-detailed case studies to future work. Nevertheless, we include a description 
of our evaluation plans and early successes.

\section{Framework Part 1: Working across the University}
\label{sec:levels}

Like others~\cite{AAAS2011}, we take academic departments as the key unit of change in a
university, because faculty are most likely to be impacted by the culture of their department and
interactions with other faculty in their department. To support departmental change, we propose
a framework that focuses on three levels of the university: faculty, department,
and administration (see Table~\ref{tab:levels}). Each level represents a subsystem of the university
(i.e., a collection of people, structures, and norms) that can be acted upon by a change process,
with different types of change processes being appropriate for different levels. This framework
does not include other actors (e.g., students, postdocs, staff) despite their important roles
in the university, because we focus on actors most likely to influence department culture.

Although we distinguish between these levels to help focus change efforts, the boundaries
between them can at times be blurry (e.g., some individuals serve dual roles as both faculty and
administrators). Moreover, these levels are closely interrelated. For instance:
\textit{administrative} measures of teaching effectiveness influence \textit{faculty} classroom
practices~\cite{Henderson2014}; \textit{department} chairs influence how educational resources are
allocated by \textit{administrators}; and \textit{departmental} norms around evaluating teaching for
tenure impact \textit{faculty} educational practices. Given the interactions between levels,
change efforts should focus on all of these levels, not a single level in isolation.

Despite the interrelations between levels, most prior change efforts have focused on a
single level. In a 191-article meta-analysis of change in STEM education~\cite{Henderson2011},
the authors found that efforts fit cleanly into a four-category typology. The categories (with their
observed prevalence) are: (1) disseminating curriculum and pedagogy (30.4\%), (2) developing
reflective teachers (33.5\%), (3) enacting policy (27.7\%), and (4) developing shared
vision (8.4\%). These categories align with our three levels: (1) and (2) work at the faculty level,
(3) at the administration level, and (4) at the department level.

The review concluded that efforts from categories (1) and (3) are ``clearly not effective''
in isolation despite the fact that 85.3\% of the articles analyzed fit into a single
category~\cite{Henderson2011}. Moreover, the authors concluded that promoting change
``require[s] understanding a college or university as a complex system and designing a strategy that
is compatible with this system.'' Thus, change efforts should target the university at
multiple levels and account for the interrelations between these levels.

To illustrate our framework, we will discuss one example of our activities from each level
in $\S$~\ref{sec:examples}. The activities are: (1) Departmental Action Teams (DATs; faculty level),
(2) visioning and alignment (department level), and (3) developing a Teaching Quality Framework
(administration level). DATs are designed to empower a group of faculty
in a department to achieve an educational goal of mutual interest and departmental importance;
they pay explicit attention to department culture to make meaningful changes that can be sustained
by the department. Our visioning and alignment process involves working directly
with an entire department to create a coherent vision and to establish mechanisms
for achieving that vision. At the administration level, we describe an effort to create
a Teaching Quality Framework for use in promotion and tenure decisions.

To align our activities across levels, we focus on a common set of goals for departmental change;
we seek to create functional collaborative processes~\cite{Conversant2014} for supporting
research-based, student-centered teaching~\cite{Freeman2014} and increasing equity
and diversity~\cite{PCAST2012}. These goals are embodied in a set of six core commitments
(see $\S$~\ref{sec:commitments}); we use these commitments to design our interventions and
assess progress towards achieving our goals (see $\S$~\ref{subsec:eval}). They also help us to
align activities across departments; as we discuss in $\S$~\ref{sec:examples}, our efforts attend
to the unique histories of each department and are therefore different in each. Regardless
of what goals drive a change effort, it is crucial that they be made explicit, to support
the alignment and evaluation of activities. Our particular choice of goals was driven by recent calls
for change~\cite{PCAST2012} and best practices suggested by the literature~\cite{Freeman2014}.

\section{Framework Part 2: Incorporating Multiple Perspectives of Change}
\label{sec:theories}

\TabInsights

Most higher education change efforts have been driven by implicit and sometimes contradictory change
logics, which has limited their impact~\cite{Kezar2013}. The failure to make explicit the logic
underlying a change effort can lead to activities that are insufficient or even detrimental
to achieving the desired outcomes. In contrast, the literature on institutional change in business
and government settings focuses on systematically understanding change processes through
the theoretical perspectives that underlie them~\cite{Real2005}. To increase the impact of STEM
education change efforts, it is imperative to draw from this literature.

Kezar's recent book on change in the college setting provides a detailed theoretical synthesis
of the literature on organizational change (including both successful and unsuccessful efforts)
and categorizes the logics that underlie attempts at change into six broad categories: scientific
management, evolutionary, political, social cognition, cultural, and institutional~\cite{Kezar2013}.
Each of these categories describes a \emph{perspective} that helps one attend to important aspects
of the change process at different levels.  While there exist other classifications
of the organizational change literature, we draw on this particular classification both because
it is applicable to higher education and to draw further attention to Kezar's work.
While we describe each of these six categories as a singular perspective to emphasize its relation
to the other five, we acknowledge that there are a variety of viewpoints within each perspective
(i.e. these perspectives are not singular and unified).

In the rest of this section, we briefly describe the key features of each of the six perspectives
(summarized in Table~\ref{tab:insights}).  In addition to providing a summary of Kezar's categories
(which we believe is valuable for the PER community in and of itself), we analyze them
in the context of STEM education by drawing examples of the logic statements that underlie
actual change efforts from the STEM education literature and analyzing how these logic statements
relate to the change perspectives.  We draw these statements from two sources
that provide representative samples of STEM education change logics: an ethnographic study
of the third annual forum of the National Institute for Science Education~\cite{Seymour2002}
and a systematic review of STEM education change efforts related to the four categories discussed
in $\S$~\ref{sec:levels}~\cite{Borrego2014, Henderson2011}.

\subsection{Scientific Management Perspective}

The scientific management perspective~\cite{Vandeven1995} emphasize the use of incentives and rewards
to change behavior. Change efforts are seen as guided by organizational leaders who are responsible
for aligning goals, setting expectations, modeling behavior, managing communication,
issuing rewards, and providing feedback and evaluation~\cite{Brill1997,Carnall2007,Huber1993,
Peterson1997}. This perspective assumes that an organization will respond to leaders' guidance
in a purposeful, adaptive manner~\cite{Carnall2007,Peterson1995,Vandeven1995}, and that
all organizations should respond in a similar way to similar activities. Change efforts
in this category often focus on the ongoing diagnosis of problems and generation
of solutions~\cite{Golembiewski1989,Goodman1982} or on the modification of organizational structure
to create change~\cite{Guskin1996}.

Scientific management is exemplified by the logic that:
\begin{quote}
``The fastest and most enduring way to promote a renewed emphasis on teaching in the service
of learning in higher education is to restructure the faculty rewards
system''~\cite[p.~97]{Seymour2002}.
\end{quote}
This example emphasizes a top-down approach: if leaders provide the right incentives,
faculty will change.  While reward structures are important, purely top-down approaches
are generally not effective in higher education~\cite{Oakes2005,Fullan1999}
(e.g., because universities have a more diffuse organizational structure than typical corporations).

Efforts to train instructors in the correct use of new educational innovations
through structured activities rely on scientific management. Their underlying logic is that:
\begin{quote}
``STEM undergraduate instruction will be changed by developing research based instructional
``best practices'' and training instructors to use them.  Instructors must use these practices
with fidelity to the established standard''~\cite[p.~230]{Borrego2014}.
\end{quote}
Under this logic, instructors are not the primary judge of the value or effectiveness of the
``best practices'' they are implementing.  Instead, educational experts design
the practices, the intervention by which the instructors will adopt the practices, and the metrics
by which this adoption will be judged.  These interventions are generally
ineffective~\cite{Henderson2011,Seymour2002} because they fail to address the internal and external
pressures (e.g., beliefs about ``good'' teaching or tenure guidelines) that heavily influence
what the instructor does in the classroom.

The scientific management perspective provides an array of practical strategies to generate change,
such as changing incentive structures. However, this perspective makes assumes that are 
often invalid, such as a strong institutional hierarchy, completely rational actors, 
and organizational structures for which context is unimportant.

Drawing on scientific management, our efforts focus on providing greater incentives
for innovative teaching, including the revision of promotion and tenure guidelines around teaching.

\subsection{Evolutionary Perspective}

The evolutionary perspective~\cite{Morgan1986} highlights the power of external factors
(usually economic) to drive change and the need for an organization to be able to respond
to unplanned and unavoidable changes. This perspective deemphasizes human agency in initiating
change~\cite{Hrebiniak1985}; instead, the role of leaders is to manage and respond
to inevitable changes. Organizations can prepare themselves for change through
proactive monitoring of and rapid responses to external factors~\cite{Cameron1991},
creating nimble infrastructure, and not allowing any part of the organization to weaken
(since one never knows when an external factor will increase the importance of a particular part).
The evolutionary perspective also emphasizes the complexity of organizations and interrelations
between parts.

The logic that:
\begin{quote}
``Attempts to alter single elements in a complex social system will not be effective: each element
must be aligned with the others for system changes to prevail''~\cite[p.~96]{Seymour2002},
\end{quote}
draws from an evolutionary perspective by acknowledging the interdependent, complex nature
of universities and the fact that coherence is generally not built into the university system
by default (e.g., at the level of course design, there is no mechanism for assuring that
course goals, assessments, and pedagogical techniques are aligned). Thus, if there is
no deliberately-imposed coherence, university structures will evolve towards incoherence,
especially if there are other factors to encourage that shift (e.g., financial pressures
or the rise of online education).

Complexity leadership efforts explicitly acknowledge the complex, interrelated nature
of organizations and the difficulty in controlling such complex systems.  Their underlying logic
is that:
\begin{quote}
``STEM undergraduate instruction is governed by a complex system. Innovation
will occur through the collective action of self-organizing groups within the system.
This collective action can be stimulated, but not controlled''~\cite[p.~241]{Borrego2014}.
\end{quote}
To stimulate change, complexity leadership requires change agents to disrupt existing patterns,
encourage novelty, and act as sense makers; yet, the outcome of the change is largely
out of the control of the change agent.

Systemic thinking and the acknowledgement of the importance of external factors are strengths
of the evolutionary perspective that are highly relevant to higher education. However,
the evolutionary perspective is weakened by its sometimes unfounded assumption that individuals
cannot do much to impact the change process.

A systems approach to institutional change, which emphasizes that efforts must be aligned
at multiple levels, is at the core of our change framework. Moreover, our efforts focus
on creating structures that are flexible, so that they can be sustained and adapted over time
in response to changing external forces.  Both of these ideas draw from the evolutionary
perspective.

\subsection{Social Cognition Perspective}

The social cognition perspective~\cite{Kezar2001,Argyris1999,Schon1983,Morgan1986,Weick1995}
emphasizes the impact of the thought processes of individuals on change
initiatives~\cite{Morgan1986,Weick1995}. This perspective assumes that resistance to change
often results from a lack of understanding of a change process or its implications for one's work,
not outright disagreement with the change itself. In this situation, a change agent can help members
of their organization to change their thinking, a task which is complicated by the fact that
different people interpret the same environment differently~\cite{Cameron1988}. Hence, change agents
need to be able to see the institution through a variety of lenses to help others adopt
unfamiliar worldviews.

Many social cognition change efforts involve helping individuals make explicit
the unconscious aspects of their worldview (referred to as ``\emph{mental maps}'') and
confront prior beliefs with new information (i.e. using cognitive dissonance to encourage
learning~\cite{Argyris1999}). This sensemaking is facilitated by encouraging
interactions between individuals to help ``synchronize'' mental maps and by providing
professional development aimed at reexamining assumptions. Change agents can also support
organizational learning by creating data teams and enhancing the infrastructure
for collecting and interpreting institutional data. From the social cognition perspective,
it is also possible for change to occur spontaneously if members of an organization
notice a dissonance without outside intervention and then move to eliminate it.

The social cognition perspective is associated with the concepts of ``single-loop learning''
(or ``first order'' change) and ``double-loop learning'' (or ``second order''
change)~\cite{Argyris1982,Argyris1999,Schon1983}.
The former is learning/change that improves what the organization already does while retaining
existing organizational norms, goals, and structures. The latter is learning/change that challenges
existing organizational structures to arrive at innovative solutions to problems
that arise due to inconsistencies between organizational beliefs, actions, and consequences.
Second order change is much more difficult to enact than first order change because the thought
processes that lead to second order change can be threatening or embarrassing to individuals
or to the organization. Hence, changes that arise ``naturally'' are generally first order.

The logic statement that:

\begin{quote}
``Good ideas, supported by convincing evidence of efficacy, will spread ``naturally''.
On learning about the success of particular initiatives, others will become convinced enough
to try them''~\cite[p.~92]{Seymour2002},
\end{quote}
is derived from social cognition because it assumes that change results from individual
learning. However, the changes required for the systemic use of research-based teaching practices
are frequently second order because they challenge existing norms and structures; thus,
they are unlikely to occur naturally.

The underlying logic of learning organizations is that:
\begin{quote}
``Innovation in higher education STEM instruction will occur through informal communities
of practice within formal organizations in which individuals develop new organizational knowledge
through sharing implicit knowledge about their teaching. Leaders cultivate conditions for both
formal and informal communities to form and thrive''~\cite[p.~240]{Borrego2014}.
\end{quote}
Thus, in a learning organization, all parts of an organization (not just the top management
or a group of experts) continually develop and evaluate new ideas that lead to changes
in the organization.  This knowledge generation occurs when individuals make their
mental maps explicit and public, leading to second order change.

The social cognition perspective accounts for the complicated nature of human beings
in the change process and the critical role of individual knowledge. They also distinguish
between first and second order changes.  However, social cognition is incomplete because it focuses
on learning via rational and traditional forms of evidence, without sufficient attention
to learning through social and emotional (``irrational'') means.

Our efforts draw heavily from social cognition. In the facilitation of DATs, we pay close attention 
to the underlying reasoning (mental maps) of the faculty involved in the process
and its implications for our activities. Similarly, our visioning and alignment process uses surveys
and interviews to elicit the underlying reasoning used by the faculty. By understanding the way that
faculty members think about education and change, we are better able to develop a process
that aligns with how they actually reason, rather than relying on some idealized notion
of faculty thinking.

\subsection{Cultural Perspective}

The cultural perspective~\cite{Schein2010,Morgan1986,Peterson1997} emphasizes the importance
of context, values, beliefs, irrationality, fluidity, and complexity in the change
process~\cite{Collins1998, Neumann1993}. This perspective assumes that organizational change occurs
as a result of cultural change, that is, a change in the values, beliefs, myths, and rituals
of the organization. To succeed, change agents must understand the values that underlie
an organization and align their messages about change with existing or aspirational values.
They can also try to shift values by altering mission statements or using existing symbols
or rituals in new ways. Changes in culture are generally believed to be slow, unpredictable,
and ongoing processes that occur ``below the surface.''  As such, they can occur without direct
guidance, and the implicit nature of culture means that change agents often overlook its importance.

Like social cognition, the cultural perspective assumes that different individuals
in an organization hold differing views as to the nature of the organization's culture.
They also assume that change can be beneficial or harmful and can result in unintended
consequences~\cite{Smircich1983}. Because culture is such a deeply-rooted part of human experience,
cultural is particularly relevant to second order change processes.

The cultural perspective informs the change logic that:
\begin{quote}
``Finding the means to leverage relevant shifts in departmental values and practices
is the critical factor in determining whether the efforts of faculty---as individuals
and groups---and of their institutions, will be able to improve the quality of [STEM] education,
or achieve the wider goal of science-for-all''~\cite[p.~96]{Seymour2002}.
\end{quote}
This logic focuses on change as being driven by shifts in values (although it ignores
other components of culture, like symbols and rituals).  This logic takes the department
as the key unit of change in a university~\cite{AAAS2011}, because faculty are impacted
most strongly by the culture of their department as compared to the cultures of other parts
of the university or the institution as a whole.

Our efforts are rooted in a need to understand existing departmental and institutional
culture and the history of practices and relations within the departments.
In both our DATs and our visioning and alignment process we have conducted a number
of interviews to understand the relevant departmental cultures. Moreover,
our efforts take advantage of the cultural shifts towards improved STEM education that
have been generated by prior educational change efforts on the target campus~\cite{SEI2014}.

\subsection{Political  Perspective}

The political perspective~\cite{Kotter1985,Vandeven1995,Morgan1986} emphasizes the importance
of collective action as a tool for change. Change agents can use agenda setting, coalition building,
mapping power structures, and negotiating to achieve their goals~\cite{Kotter1985}.
Philosophically, this perspective draws from the Hegelian-Marxian viewpoint~\cite{Marx1867}
that ideas (norms, values, beliefs) and their opposites are always present in an organization,
and it is when these are brought into conflict (often due to resource constraints) that rapid,
second order, radical change occurs~\cite{Gersick1991,Morgan1986}. These rapid changes punctuate
long periods of slow, evolutionary change during which most members of the organization
are disengaged from the potential conflict~\cite{Baldridge1977,Conrad1978}. The political perspective
emphasizes that change can be erratic, irrational, and potentially regressive. While changes
may benefit only certain groups, empowerment approaches encourage changes that mutually benefit
everyone involved~\cite{Astin1991,Bensimon1993}.

A logic statement informed by the political perspective is that:
\begin{quote}
``Change can be built from small local beginnings, first by provoking and maintaining
conversations that lead to local collaboration; then by making connections with collaborators
on the same or other campuses~\cite[p.~96]{Seymour2002}.
\end{quote}
This statement suggests that agenda setting and coalition building can sow the seeds of change.
However, it does not explicitly address the existence of opposing camps that may come
into conflict with this coalition.  If the change agent fails to deal with these opposing camps
effectively, then the change effort will be in jeopardy.

The political perspective emphasizes that individuals positioned at all levels of an
organization can effect change.  In the context of a university, coalition building and
mapping power structures are particularly important, especially for change agents outside of the
traditional administrative power structure.  Nevertheless, the political perspective tends to ignore
important ideas from social cognition (e.g., that resistance could be due
to misunderstanding rather than competing interests). None of the efforts discussed by Borrego and
colleagues~\cite{Borrego2014} have a political perspective as a core underlying logic.

Our own efforts focus on how to align the goals and agendas of various actors to achieve
our goals of institutional transformation. For instance, we have aligned our activities with
the upper administrations's charge to the target university to improve student retention. 
Within our DATs, we have paid explicit attention to the composition of team members,
so as to include members with strategic influence within the department.

\subsection{Institutional Perspective}

The institutional perspective~\cite{Leicht2008,Powell1991,Slaughter2004} blends ideas from other
perspectives, but is uniquely characterized by the attention it pays to the relationship
between a target institution (e.g., a college or university) with the network
of other institutions that exert influence over it (e.g., accreditation agencies, professional
societies, and legislatures). This perspective places emphasis on the pressure to change exerted
by this external network on an institution as it tries to maintain legitimacy,
while also acknowledging that institutions can have significant ``inertia'' to resist change,
especially those with long-standing missions and identities~\cite{Powell1991}. Isomorphism,
the tendency of similar institutions to converge in their missions over time, is a central concept
in institutional perspectives~\cite{Morphew2009}. Like the evolutionary perspective,
the institutional perspective emphasizes the need to understand the impact of external institutions
over which one typically has little direct control.

The underlying logic that:
\begin{quote}
``The time for development, implementation, and testing that agency grants provide,
plus the prestige of such awards, will increase the chances that innovation will take root
in the host institutions beyond the end of funding''~\cite[p.~100]{Seymour2002},
\end{quote}
couples the potential sustainability of an educational innovation with the support provided
by funding agencies through the awarding of grants.  While grants carry institutional prestige,
this logic statement ignores other external institutional factors that may work against
the sustainability of an education innovation.  Moreover, there is no guarantee that the university
will continue to support the innovation once external funding runs out.

In higher education, the process by which a university conducts an external review of one
of its programs, typically to satisfy accreditation agencies, is an example of quality assurance.
This process is central to the following change logic:
\begin{quote}
``STEM undergraduate instruction will be changed by requiring institutions (colleges, schools,
departments, and degree programs) to collect evidence demonstrating their success in undergraduate
instruction. What gets measured is what gets improved.~\cite[p.~235]{Borrego2014}.
\end{quote}
Here ``quality'' is defined by an external institution and may or may not align with the best
interests of students, faculty, or staff.  Additionally, one can imagine a future in which other
institutions require similar forms of quality control (e.g., the federal government requiring that
universities meet certain standards in order to receive student aid money).  Hence, it is
in a university's best interests to have as much say as possible in the process by which
the external institution decides what is to be measured.

In our own efforts, we draw on the institutional perspective by leveraging the prestige
of our funding source to promote the legitimacy of our efforts. We have also aligned
our efforts with funding and political shifts within the target university institution 
(e.g., aligning our work with the student success initiative, which is largely externally-driven).

\subsection{The Need for Multiple Perspectives}

\TabExamples

Each of these perspective provides key insights into a change process but is also
limited in its focus. Thus, change efforts are most likely to succeed when they draw
from all six perspectives~\cite{Kezar2013}.  However, each of the logics described above tends
to draw from a limited subset of these perspectives (a different subset for each logic); they do not
work across all of the perspectives holistically.   While a formal analysis of all of the change
logics described in systematic literature reviews~\cite{Seymour2002, Henderson2011} is beyond
the scope of this paper, the sample provided above is generally representative of the complete set.
Our team has formally analyzed all of the logics described in the reviews, but only presented
a subset of them here, due to space constraints.

Consideration of all six change perspectives can shed light on the reasons why
some change efforts may not be successful. For instance, efforts focused on disseminating curriculum
and enacting policy are generally ineffective in isolation~\cite{Henderson2011}.
Dissemination efforts often focus on changing practices, but fail to account for underlying beliefs
(\textit{social cognition} perspective), departmental culture (\textit{cultural} perspective),
and institutional incentive structures (\textit{scientific management} and \textit{institutional}
perspectives). Similarly, policy efforts often ignore underlying beliefs and departmental culture
(\textit{social cognition} and \emph{cultural} perspectives, respectively). Taken together,
this analysis supports the claim that \textit{change efforts should draw from all six change
perspectives}.  Given the relative dearth of examples of efforts in the STEM education
literature that do so, we describe our activities to illustrate how these perspectives can be used
to guide and inform change.

\section{Illustrating the Framework}
\label{sec:examples}

To illustrate our change framework, we provide an example of our change efforts
at each level of the university, emphasizing how the change perspectives from $\S$~\ref{sec:theories}
informed our activities (summarized in Table~\ref{tab:examples}). We describe our activities
in two departments: the Runes Department and the Charms Department (actual names
redacted for confidentiality), both of which participated in the Science Education Initiative
(SEI)~\cite{SEI2014,Chasteen2011}.  SEI was an initiative to support departmental
transformation through the adoption of learning goals and practices that support those goals.
A key resource provided to departments though the SEI was postdoc-level Science Teaching Fellows
(STFs) who helped individual faculty and departments develop and adopt effective practices.

Because all academic institutions are unique, change efforts necessarily depend
on local context; in fact, our activities vary among departments.  We do not claim
that our particular activities can be directly exported to another context. Rather,
we describe our activities to show how change perspectives can be used to guide a holistic change
effort. Because our purpose is not to evaluate our activities themselves, we describe them only
as much as necessary to ground our framework.  Moreover, because our efforts are ongoing,
we do not claim that our activities have ``resulted'' in change.  Nevertheless, we discuss
principles for generalizing from our experiences in $\S$~\ref{subsec:generalize} and 
mechanisms for assessing the success of our change efforts in $\S$~\ref{subsec:eval}.

\subsection{The Faculty Level}
\label{subsec:DAT}

An example of our work at the faculty level is the creation of Departmental Action Teams (DATs).
Like faculty learning communities (FLC; \cite{Ortquist2009}), DATs consist of self-selected faculty
who have agency to choose an educational issue they will address. DATs differ from most FLCs because
they focus on a common, shared goal in a single department rather than individual projects
in multiple departments. The aim of a DAT is to create lasting department structures that address
this shared goal in a sustainable way. Thus, DATs focus on goals that are relevant
to their department broadly (i.e. not just transforming a single course).

We formed a DAT in the Runes Department in September 2014.  The Runes Department
was created out of components of two other departments about a decade ago, and one of its initial
challenges was defining a curriculum for its majors.  Soon after its formation, Runes
became involved with the SEI, which helped shape the department's emerging educational culture.
SEI involvement led to the development of learning goals and transformed pedagogy in most
of the department's required courses, the involvement of many of the department's full-time
(non-tenure track) instructors in scholarly teaching (e.g., publishing in education journals),
and the perception by some faculty, including departmental leaders, that their department is
on the forefront of educational innovation for Runes Departments nationwide.
We elicited this information through interviews with 9 of the department's 31 faculty members
(2 of whom joined the DAT).

In forming the DAT, we leveraged the prestige of our funding source, as suggested by
the \emph{institutional}
perspective, to gain access to departmental leaders.  Moreover. we built coalitions and leveraged
existing departmental power structures, as suggested by the \emph{political} perspective.
We met individually with a subset of the interviewed faculty to get their input (and buy-in)
into the idea of a DAT in their department.  We then secured the sanction of the department chair
and teaching committee to form the DAT, which was announced at a faculty meeting.  As recommended
by the \emph{cultural} perspective, we tied into existing culture by framing the DAT
as a continuation of SEI's progress, which resonated with faculty, particularly those who feared
that the loss of formal SEI support would lead to backsliding in this progress. Recently,
the department decided to write about the DAT in its department newsletter, indicating ongoing
interest within the department. Additionally, in alignment with \emph{scientific management},
the chair provided all DAT members with service credit and one instructor with a course buyout.

The Runes DAT consists of five faculty participants (one tenured professor, three
full-time instructors, and one retired instructor) and two external facilitators.  This group met
for 16 hour-long meetings over the 2014-2015 academic year.  The initial plan was for the DAT to end
after one year, but the participants have expressed their desire to continue the DAT throughout
the next academic year as well.  The DAT aims to create coherence across the Runes
curriculum by: (1) creating greater awareness and use of existing learning goals, (2) facilitating
communication between faculty across courses, and (3) weaving non-content goals (e.g., experimental
design and scientific communication) throughout the curriculum in an integrated fashion.

The DAT is co-facilitated by two members of our project team.
The facilitators focus on creating an inclusive, collaborative, data-driven environment
(in alignment with our core commitments). At the first DAT meeting, the group jointly constructed
a set of outcomes that they desired for Runes majors, which helped the group articulate
their goals for the department. Additionally, the facilitators have provided the DAT with data
from the university's institutional research office to check the accuracy of anecdotal claims about
Runes students (e.g., that many students transfer credit for Runes courses
taken at other institutions). These activities align with the \emph{social cognition} perspective
because they help DAT participants revise their unconscious views and prior beliefs through
data analysis and group learning.

The facilitators also help the DAT participants align their work with the change perspectives
discussed in $\S$~\ref{sec:theories}. To create coherence across the curriculum, the DAT proposed
the creation of three new curriculum coordinator positions, each associated with a different subset
of the core Runes courses.  The coordinators will facilitate communication
among the faculty teaching these courses, maintain continuity in learning goals
across the curriculum, assess student learning outcomes, and organize professional development
activities around teaching.  In effect, these coordinators will prepare the department to adapt
to unforseen curricular changes, in alignment with the \emph{evolutionary} perspective.
Additionally, these activities will lead to departmental change by influencing
the way that faculty see their role as instructors and the relationships among their courses;
this is aligned the \emph{social cognition} perspective.  Because these changes could
be perceived as threatening by some faculty, the \emph{cultural} perspective suggests that the
coordinators will have to align their activities with established departmental values and norms
to mitigate the chance of rejection by the department.

The \emph{political} perspective influenced how these coordinator positions were proposed.
Rather than starting at a faculty meeting, the DAT participants met directly
with the department's chair and teaching committee, and the committee allocated
three course buyouts to allow three instructors to fill these positions.  The support of the
department leadership makes it less likely that skeptical faculty members will be able to derail
the implementation of the coordinator plan.  Nevertheless, both the DAT participants and the
new coordinators will need to leverage their personal connections to build a coalition of supporters
among faculty who have not been part of the DAT process. This base of support will make it easier
for the DAT to implement changes to shift the departmental culture around teaching and learning.

\subsection{The Department Level}
\label{subsec:VAP}

An example of our work at the department level involves the implementation of a visioning and
alignment process at the scale of an entire department.  Large-scale cultural change processes
have been implemented in business organizations for decades~\cite{Real2005}, but they have not been
systematically applied in higher education. To do so, we adapt the \emph{cycle of value}
approach~\cite{Conversant2014}, which consists of three phases implemented iteratively:
(1) align, (2) act, and (3) adjust; this approach has demonstrated success in other ``knowledge
intensive'' organizations similar to academic departments. In alignment with successful change
efforts~\cite{Nutt1997} our activities will: (A) help the department develop a clear vision of the
end state they wish to achieve, (B) focus discussions on outcomes rather than problems,
and (C) emphasize the value of collective goods (e.g., learning goals guiding the major)
over individual rights (e.g., faculty ``ownership'' of courses).

We are presently adapting this approach in the Charms department, a relatively low-conflict
department with a high degree of commitment to education.  We began our change process
by contacting the departmental leadership (as suggested by the \emph{political} perspective),
including the chair, the executive committee, and a senior faculty member who is the director
of Charms' SEI efforts.  This faculty member has become a champion for Charms'
involvement with our initiative; the existence of such a champion makes it more likely that his
colleagues will engage productively with the change process~\cite{Kotter1996}.

Because nearly all decisions in the Charms Department are made democratically,
we built support for our change effort by presenting it at a faculty meeting and allowing
the Charms faculty several weeks to discuss possible involvement with us (informed by the
\emph{cultural} perspective). Over time, the department began to see our change process
as a means to continue work they had begun through the SEI: the development of department-wide
learning goals and a structure to sustain their use. Continuing this process would allow Charms
to continue to be seen as an educational leader on the target campus and nationally, thus leveraging
the department's own self-perception (informed by the \emph{cultural} perspective),
the interconnected nature of the university system (informed by the \emph{evolutionary} perspective),
and the prestige of our funding source (informed by the \emph{institutional} perspective).
Ultimately, the department voted unanimously to engage in the change process.

Since securing departmental support, we have begun the \emph{alignment} phase of our change process,
which begins by understanding the mental maps that individuals within the department use to reason
about education. By uncovering and shifting these maps, we aim to foster second order changes
in thinking (informed by the \emph{social cognition} perspective). To elicit these mental maps,
we administered a departmental survey to probe the alignment between faculty aspirations
for their department and their perceptions of its current state, and we conducted individual
interviews with faculty that probed their educational goals, understanding of change,
and perception of departmental climate, governance, and decision-making. In both cases,
we received responses from 26 of 36 faculty (72\%).  Using this information, we created a detailed
picture of the current state of the department (e.g., knowledge, norms, and relationships),
which we will use to guide our change process (in alignment with the \emph{social cognition},
\emph{cultural}, and \emph{political} perspectives). We also reported our findings back
to the department to facilitate their own sensemaking.

Additionally, we shared our core commitments with the department as a starting point
for creating a shared vision (see $\S$~\ref{sec:commitments}).  In essence, our core commitments
lay out a basic structure for what the change process might achieve, but the department must
build upon and interpret them to create a vision that is consistent with their existing values
and norms (consistent with the \emph{cultural} perspective).  By the end of the 2015 spring semester,
we will have facilitated a 2-hour meeting and a day-long retreat with Charms to develop
a shared vision, uncover and revise unhelpful assumptions about education, and create 30-, 90-,
and 180-day action plans.

The \emph{action} phase involves moving towards the shared vision by enacting the action plans
and building capacity as needed. Through our interviews, we have uncovered that the department
will need to find ways to overcome time constraints and bolster mechanisms for communication
and collaboration to build capacity for this change process. This will require redistributing
existing resources or seeking out additional support where it is required (aligned with the
\emph{scientific management} perspective). We will also aim to create early wins,
to help increase faculty buy in (aligned with \textit{political}, \textit{cultural},
and \textit{social cognition} perspectives). To make this process sustainable, the department
will need to integrate teaching and learning goals systematically with research and other
departmental goals. This step will require revising mental maps of how some faculty view teaching
(aligned with the \emph{social cognition} perspective): not simply as an ``add on'' but as equally
important as other departmental activities and a meaningful part of scholarly practice.

The \emph{adjustment} phase focuses on sustaining this process in the long term.
During adjustment, the department will assess the success of plans implemented in earlier phases
and use the insights gained to adjust their goals and generate new action plans (in effect,
circling back to the alignment phase).  This will involve creating mechanisms to reward success
and understand failures, thereby reinforcing the faculty's collective dedication
to the shared vision (in alignment with \emph{scientific management}, \emph{social cognition},
and \emph{political} perspectives).  Ultimately, these mechanisms can increase the department's
capacity with respect to research and service, in addition to teaching, if they become deeply
embedded in departmental governance and decision-making processes (as suggested by
the \emph{cultural} perspective). We plan to engage in the first adjustment phase next fall.

We anticipate that the change process, facilitated by our project team, will last between one
and two years. Additionally, our team has been in contact with senior administrators at the target
university to help secure financial and other resources as required.  After the first 
adjustment phase is completed, the department will once again align its objectives and take 
new actions to achieve them. As the department continues through these cycles, it will continue
to achieve greater alignment and coherence on its way to achieving its shared vision.

\subsection{The Administration Level}
\label{subsec:TQF}

At a research university, investment in effective teaching is often viewed as conflicting
with research productivity, which is the primary driver of career advancement for faculty.
Accordingly, an example of our work at the administrative level focuses on aligning incentive
structures with innovative, student-centered learning to better reward teaching excellence
(in accordance with the \emph{scientific management} perspective).  Our major effort in this context
is the development of a Teaching Quality Framework that will be used by departments at the target
university for faculty evaluation in tenure and promotion decisions.  Such a framework would 
clarify what it means to be excellent in teaching on that campus, thus encouraging faculty to pursue
a more teaching-focused route to tenure.

While there are some precedents for the creation of such a framework on other campuses,
one significant barrier that we face is that our team holds limited administrative power 
on the target campus.  Thus, we must work within the existing institutional structures to change 
the ways of thinking among faculty and administrators (drawing on the \emph{social cognition}
perspective) to make it possible for a Teaching Quality Framework to be created, accepted, 
and interpreted in a meaningful way. Our approach focuses on: (1) aligning with external initiatives
and organizations to promote a local focus on teaching excellence (drawing on the
\emph{institutional} perspective), (2) building coalitions with key stakeholders on the target campus
to influence policy and messaging (drawing on the \emph{political} perspective), and 
(3) leveraging existing campus initiatives in our messaging (drawing on the \emph{cultural}
perspective).  As suggested by the \emph{evolutionary} perspective, we have been flexible 
and opportunistic in choosing which existing groups and initiatives to work with 
on the target campus.

Drawing on the \emph{institutional} perspective, our team has used the prestige of our funding source
to meet with key administrators to stress the urgency and timeliness of our efforts.
We have also convinced administrators to support their campus in joining the Bay View Alliance (BVA),
``a consortium of research universities carrying out applied research on the leadership
of cultural change for increasing the adoption of improved teaching methods
at universities''~\cite{BVA2014}.  The BVA researches areas such as: introductory course
transformation, cross-disciplinary intellectual skills, and data-driven decision making.
By connecting the target campus to the BVA, we aim to increase campus leaders' exposure to the ideas
of research-based teaching.

Simultaneously, we are working with important groups on the target campus (aligned with 
the \emph{political} perspective) to make small policy changes that lay the groundwork for 
a Teaching Quality Framework. For example, we have worked with the target university's faculty senate
to shift the nomination requirements for its campus-wide teaching excellence award to require
evidence of teaching effectiveness, of student development and engagement, and of contributions
to the scholarship of teaching and learning.  As suggested by the \emph{social cognition}
perspective, the redefinition of these criteria is a tool to help shift how those nominating 
their colleagues for this award understand teaching excellence, so that when a larger 
Teaching Quality Framework is created, there will be less faculty resistance due to misunderstanding
the framework's meaning.

In alignment with the \emph{political} perspective, our team is leveraging a persistence taskforce
that reports to the Provost on the target campus.  This taskforce was created in response to calls
from the campus's senior administration to improve the low retention of undergraduate
students. One of the members of our project team accepted an appointment to the taskforce,
and as a result of his participation, the taskforce issued a recommendation to the Provost
to enhance the prestige, respect, and reward structure for excellence in scholarly teaching.
In alignment with the \emph{social cognition} perspective, he also worked with this group
to reframe the conversation as one of \emph{student success} rather than \emph{retention},
thus helping to change the administrators' way of thinking about the problem to one that better
connects with teaching excellence and university culture.

With this groundwork laid, our team is working with the faculty senate and senior administrators 
to create a faculty taskforce charged with creating and implementing a Teaching Quality Framework.
The purpose of this effort is not to frame teaching as opposed to research, but rather
to frame them as mutually supportive endeavors, shifting the value structure on campus (aligned with
the \textit{cultural} perspective).  Populating the taskforce with influential and respected
faculty members will help increase faculty buy-in and make it more likely that such a framework
would actually be adopted once it is created (in alignment with the \emph{political} perspective).
Our project team has gathered resources for developing such a framework and will seed
these resources within the taskforce to support the framework's development. Moreover,
we invited a national leader on transforming promotion and tenure to advise our team.
In addition to meeting with us, he met with key stakeholders on the target campus to further
this initiative. Our aim is for this taskforce to be constituted over the fall semester so that
it may begin its work in the spring.

\subsection{Synergies Across Levels}
\label{subsec:synergies}

Each of the above efforts (and others not described in this paper) is aimed at changing the culture
of educational practices on the target campus to achieve greater alignment with our core commitments.
Thus, all of our activities are focused on a common objective. Beyond alignment, our efforts
are synergistic. For instance, our Runes DAT has facilitated structural changes
in the department, which will result in the creation of department-level learning goals
and tighter integration between courses. This is a precursor to the whole-department visioning
and alignment process that we are currently conducting with the Charms department. In this way,
the Runes DAT could help to ready the department to engage in its own large-scale
visioning process.  Thus, we target our change efforts to meet a department where it is at,
allowing it to engage in the change for which it is ready a that moment while simultaneously
growing in the capacity for more extensive change in the future.

Our faculty- and department-level work also has synergies with our administrative-level work.
For instance, the Charms department desires to be a leader on campus in its educational mission,
and it has recently succeeded in tenuring a faculty member based on both research and teaching
excellence.  In many ways, this makes Charms an ideal department for early adoption and testing
of a Teaching Quality Framework.  At the same time, administrative support for teaching
excellence, as signaled by the existence of the framework, will ease potential fears that Charms
faculty may have about the potential negative repercussions on their careers of focusing
to heavily on their teaching.

\subsection{Generalizing Beyond the Target Campus}
\label{subsec:generalize}

To effect change, one's efforts must be aligned with the existing cultures, ways of thinking,
and political structures of a particular university and its departments. Hence, one must assess
a department's ``readiness for change''~\cite{Toma2010}. However, it is not a matter of
\emph{whether or not} a department is ready for change, but rather of \emph{what type} of change
the department is ready for. In the Runes department, we felt that a DAT would be
the most productive tool for building on existing efforts given the high level of involvement
from a subset of full-time instructors and faculty in the department's SEI efforts. In contrast,
existing democratic structures and the ongoing development of department-wide learning goals led us
to engage in a full-scale departmental process in Charms.

As much as possible, our efforts leverage existing campus resources. As described above,
we have built on the SEI's impact in our two example departments.  We also acknowledge the existence
of other factors on the target campus that worked in our favor, such as the robust discipline-based
education research community and a Learning Assistant program that is well-supported
by the administration. Additionally, one of our team members is an expert on organizational
change and another is politically well-connected on the target campus.  These preexisting
conditions help to define our initial strategies, expectations, targets for change, and the types
of activities we use in our change process.  Other change agents in other contexts will have
a different set of preexisting conditions and expertise and will therefore need to assess their
strengths and weaknesses to determine where to start in their change process.  For example, they
may seek out professional development opportunities to strengthen their knowledge of institutional
change or networking opportunities to strengthen political connections on their campus.  Thus, while
the specific activities highlighted in this paper may not work in all contexts, the core ideas
in our framework (i.e., working across institutional levels and designing activities based
on the six change perspectives) are broadly applicable.

\subsection{Evaluation}
\label{subsec:eval}

While evaluating the impact of our ongoing efforts is beyond the scope of this paper,
we briefly outline our evaluation methods. Our evaluations focus both on the products
of our activities (i.e. actual changes in structures and policies) and changes in the cultural
beliefs and practices of individuals we work with. To assess our efforts, we draw on three types
of data: (1) surveys and interviews of individuals, (2) observations of group activities
(e.g, DAT meetings, retreats, or taskforces), and (3) artifacts that result from these activities
(e.g., reports, policy statements, vision statements, or new departmental structures).
To assess the impact of the project over time, we will revisit outcomes in the various departments
over the next several years; this is especially important because of evidence that educational
transformations are not always sustained~\cite{ChasteenInPress}.

At the departmental level, we will use pre/post measures to look for changes in culture.
As discussed above, we conducted interviews and surveys in Charms to begin the change process.
This ``pre-test'' provides evidence of faculty perceptions of the current departmental culture
and alignment with our core commitments (see $\S$~\ref{sec:commitments}).  We will administer
``post-test'' surveys and interviews towards the end of the change process to measure cultural
shifts.  We will also use artifacts like vision statements and observations of working meetings
to assess shifts in how faculty talk about education, make decisions, focus on outcomes versus
problems, and so on; taken together, these measures of culture will indicate the degree of alignment
with our core commitments.  We expect that Charms' change process will involve the creation
of assessment measures by the Charms faculty themselves that they will need to assess
their own progress; we will also use these as indicators of change. Finally, we will look
at the actual structural changes made within the department (e.g., creation of goals, committees,
or collaborative processes) as indicators of success.

Because DATs are more limited in scope, we will use different measures to assess their impact.
For example, we are preparing for end-of-year interviews with the faculty
in the Runes DAT. These interviews will focus on understanding the ways in which
the DAT members perceived themselves to be change agents within the department and which aspects
of the DAT's structure and facilitation were most crucial to the success of the DAT.
We will use our records of DAT meetings and the documents produced by the DATs to triangulate
the results from our interviews. We will also look at outcomes, such as the creation of standing
coordinator positions to integrate learning goals, as a sign of success. We will continue
to evaluate the impact of these positions, and other DAT activities, over time.

At the administrative level, all of our assessment will focus on the analysis of policy statements
and structural changes that stem from our work.  We will use these to measure the level of support
for innovative education exhibited by the administration.

\section{Conclusion}
\label{sec:conclusion}

The improvement of higher education requires more than the development of new teaching strategies;
it requires systemic, cultural change. However, most approaches to change in higher education
have been plagued by a number of limitations:
\begin{enumerate}
\item They focus on changing practices at the exclusion of changing culture.
\item They ignore the complex, interrelated nature of universities, focusing
on only one part of the system.
\item They do not adequately draw from the vast organizational change literature.
\end{enumerate}

We address these limitations by introducing a framework for holistic change.
This framework encourages change agents to draw from a wide spectrum of change perspectives
to target the faculty, department, and administration of a university in a coordinated fashion,
with the ultimate goal of changing departmental culture.  We hope that other who may wish
to engage in their own change efforts will find the framework and examples that we have provided
helpful in carrying out that work.

\begin{acknowledgments}
We thank the Association of American Universities and the Helmsley Charitable Trust for funding
this work through the AAU STEM Education Initiative. We also acknowledge the important role played
by Kezar's categorization of the organizational change literature in our analysis.
\end{acknowledgments}

\bibliography{../../_Resources/references}

\appendix

\section{Core Commitments of Target Departmental Cultures}
\label{sec:commitments}

Because our overall strategy for shifting departmental culture requires coordination among
multiple activities, it is important that we have a clear vision of the culture we are trying
to create so that we can align our activities with that goal.  Our target departmental culture
is described by six core commitments; we believe that a department that embodies these commitments
will create an improved educational environment for STEM undergraduates.  We note that
other projects at other universities may have different specific goals or commitments,
but it is important for such goals to be explicitly stated to help coordinate activities
across levels.

These are the commitments as given to the Charms Department as part of their visioning
and alignment process:
\begin{enumerate}
\item[\textbf{C1}] \textbf{Students are viewed as partners in the education process:}
Students play an active role as partners in the education process, not simply as recipients of
education. Students' current understandings are seen as a resource to be built upon,
and students engage in higher-order thinking as part of their course experiences.
Students have opportunities to exercise agency and voice in their education by playing
an active role in setting outcomes and goals for their academic program.
\item[\textbf{C2}] \textbf{Educational experiences are designed around clear learning outcomes:}
Outcome thinking focuses on the end states to be achieved.  Thus, learning outcomes specify the
qualities, capacities, and understanding desired for students at the end of any given learning
experience (from an individual assignment, to a course, to the major as a whole).
The determination of appropriate outcomes is guided by the long-term developmental
needs of students as people, scholars, and professionals in their field of study.
Choices related to pedagogical practices are guided by these learning outcomes rather than
a priori preferences.
\item[\textbf{C3}] \textbf{Educational decisions are evidence-based:}
The department and its faculty use evidence as the basis for making educational decisions, with
a clear process for doing so. The department collects meaningful data about student learning
outcomes to assess whether or not students are actually meeting these outcomes. The department
regularly consults the educational research literature in its decision-making.
\item[\textbf{C4}] \textbf{Active collaboration and positive communication exist within the
    department and with external stakeholders:}
Faculty, students, and staff engage in an ongoing dialogue about education that reflects their
shared, collective responsibility towards meaningfully supporting student learning. Mechanisms
exist for identifying, understanding, and resolving conflicts among department members and with
constituent groups.  The department has informal gathering spaces that encourage discussion,
collaboration, and community building among faculty.  The department exhibits evidence-based
best practices in decision-making processes.
\item[\textbf{C5}] \textbf{The department is a ``learning organization'' that is focused on continuous improvement:}
The department uses systems thinking, seeing department functions (e.g., teaching, research, and
service) as integrated, not separate.  Improvement takes place across the departmental system,
with explicit attention to the relationships among goals, functions, and actions. The department
develops the capacities of individual members through training and team learning,
and aligns rewards and incentives with desired outcomes (including learning outcomes).
Department members reflect on their actions, are willing to revise their assumptions, and are open
to attending to events in new ways.  These practices lead to continued learning, and as a whole,
the department becomes better at learning how to learn.
\item[\textbf{C6}] \textbf{The department values inclusiveness, diversity, and difference:}
The department makes efforts to recruit, retain, and support individuals from underrepresented
groups, broadly defined. The department is mindful that its choices will affect different
populations differently and therefore acts in ways that are supportive of all communities
within the department and served by the department. The department prepares students to work
in a diverse society and works to promote a culture of inclusiveness in society.
\end{enumerate}

\end{document}